\documentclass[]{raa}
\usepackage{graphicx,times}
\usepackage{natbib}

\providecommand{\bm}[1]{\mbox{\boldmath$#1$\unboldmath}}

\begin{document}

   \title{Measuring the beaming angle of GRB 030329 by fitting the rebrightenings in its multiband afterglow}

 \volnopage{ {\bf 2010} Vol.\ {\bf 10} No. {\bf XX}, 000--000}
   \setcounter{page}{1}

   \author{Wei Deng
   \and Yong-Feng Huang
   \and Si-Wei Kong
   }

   \institute{Department of Astronomy, Nanjing University, Nanjing 210093, P. R. China; {\it hyf@nju.edu.cn}\\
   Key Laboratory of Modern Astronomy and Astrophysics (Nanjing Univ), Ministry of Education, P. R. China\\
\vs \no
   {\small Received [year] [month] [day]; accepted [year] [month] [day] }
}

\abstract{Multiple rebrightenings have been observed in the
multiband afterglow of GRB 030329. Especially, a marked and quick
rebrightening occurred at about t $\sim$ 1.2 $\times$ 10$^{5}$ s.
Energy injection from late and slow shells seems to be the best
interpretation for these rebrightenings. Usually it is assumed that
the energy is injected into the whole external shock. However, in
the case of GRB 030329, the rebrightenings are so quick that the
usual consideration fails to give a satisfactory fit to the observed
light curves. Actually, since these late/slow shells coast freely in
the wake of the external shock, they should be cold and may not
expand laterally. The energy injection then should only occur at the
central region of the external shock. Considering this effect, we
numerically re-fit the quick rebrightenings observed in GRB 030329.
By doing this, we were able to derive the beaming angle of the
energy injection process. Our result, with a relative residual of
only 5\% --- 10\% during the major rebrightening, is better than any
previous modeling. The derived energy injection angle is about
0.035. We assume that these late shells are ejected by the central
engine via the same mechanism as those early shells that produce the
prompt gamma-ray burst. The main difference is that their velocities
are much slower, so that they catch up with the external shock very
lately and manifest as the observed quick rebrightenings. If this
were true, then the derived energy injection angle can give a good
measure of the beaming angle of the prompt $\gamma$-ray emission.
Our study may hopefully provide a novel method to measure the
beaming angle of gamma-ray bursts.  }

   \authorrunning{W. Deng, Y.-F. Huang \& S.-W. Kong }
   \titlerunning{Measuring the beaming angle of GRBs}
   \maketitle

\section{Introduction}
\label{sect:intro}

GRB 030329, which is one of the brightest gamma-ray bursts (GRBs),
has a fluence of about 1.18 $\times$ 10$^{-4}$ ergs$\cdot$ cm$^{-2}$
(\citealt{ric03}; \citealt{van04}). It is also very close to us. The
redshift is z = 0.168 (\citealt{gre03}). At the same time, GRB
030329 is unambiguously confirmed to be associated with supernova
(\citealt{hjo03}; \citealt{sta03}; \citealt{mat03}). Because of
these important characteristics, this GRB has attracted a lot of
attentions. Numerous and detailed multiband afterglow observations
have been accumulated. Multiple rebrightenings were observed in the
afterglow of GRB 030329. Especially, a quick and marked
rebrightening occurred at about t $\sim$ 1.2 $\times$ 10$^{5}$ s.

Several models have been proposed to explain the quick and marked
rebrightenings. (i) Density-jump model: if the external shock
encounters a sudden density variance of the circum-burst medium, the
emission of the afterglow may be enhanced temporarily
(\citealt{wan00}; \citealt{laz02}; \citealt{nak03}; \citealt{dai03};
\citealt{tam05}). But for GRB 030329, it has been argued
that the density-jump model could not produce the observed
rebrightenings (\citealt{nak03}; \citealt{wil04}; \citealt{pir04};
\citealt{hua06}). (ii) The two-component jet model is another kind
of choice (\citealt{ber03}). But \cite{hua06} made a detailed
numerical calculation and found that although the two-component jet
model could basically reproduce the overall R-band light curve, it
was unable to explain the steep rebrightenings. (iii) \cite{wil04}
discussed a possibility that the rebrightening might be due to
a supernova component, but their model is unlikely to
produce multiple rebrightenings.

Actually, the energy-injection model seems to be the best
explanation for the steep rebrightenings (\citealt{hua06}).
\cite{gra03} have made a general analysis on this interpretation and
\cite{hua06} presented a detailed numerical study. In Huang et al.'s
study, it is interesting to note that although the fitting result
was much better than previous models, the theoretical rebrightenings
still could not be as steep as observations. This problem was
probably caused by their simple assumption that the energy carried
by the late (and initially slower) shells was injected into the
whole external shock homogeneously. In reality, it was probable that
the late/slow shells, which coast freely in the wake of the previous
external shock, should be cold and do not expand laterally, as
illustrated by \cite{gra03}. So when the late/slow shells catch up
with the external shock, the energy should only be injected into the
central region of the external shock. We call this scenario as a
localized energy-injection scenario. We conjecture that the
energy-injection angle can be derived from the rapidness of the
observed rebrightenings. Since these late/slow shells basically may
have the same origin as those early shells that produce the internal
shocks and the main GRB, we argued that this method could be used to
hint the degree of collimation in GRBs. This will be a novel way to
measure the beaming angle of GRBs, in addition to the traditional
jet-break timing method.

In this paper, we will try to reproduce the multiband afterglow
light curves of GRB 030329 numerically, based on the localized
energy-injection scenario. The structure of our paper is organized
as follows. Section 2 is a detailed description of the model. In
Section 3, we study the effects of various parameters on the
rapidness of the rebrightening. In Section 4, we simulate the
observed multiband afterglow light curves of GRB 030329 in the
framework of our model. Section 5 presents a brief discussion and
the conclusion.

\section{Localized Energy-Injection}
\label{sect:LEI}

According to the standard fireball model, when the fireball plows
through the circum-burst medium, it will produce a strong blastwave
that accelerates the swept-up electrons (\citealt{pir99};
\citealt{van00}; \citealt{mes02}; \citealt{zha04}; \citealt{pir05}).
The afterglow is produced by these accelerated electrons. In this
study, we will only consider the synchrotron radiation from
electrons, although inverse Compton scattering may also play a role
in some cases (\citealt{wei00}; \citealt{sar01}). The detailed
condition of afterglow is quite complicated. For instance, the
blastwave may evolve from highly radiative regime (the radiation
efficiency $\epsilon$ = 1) to adiabatic regime ($\epsilon$ = 0). The
blastwave will experience the transition from ultrarelativistic
phase to Newtonian phase. Meanwhile, the effect of lateral expansion
of the jet, the condition of the circum-burst medium (homogeneous or
wind-like) and the equal arrival time surface effect
(\citealt{wax97}; \citealt{sar98}; \citealt{pan98}) are also
important factors which influence the evolution of the afterglow. In
order to reproduce the complicated behavior of the afterglow of GRB
030329, all these factors should be considered. For this purpose,
numerical evaluation is more efficient than analytical method.
\cite{hua99} and \cite{hua03} have developed an effective method
which can easily take into account all the complex conditions
addressed above. Recently, \cite{vane10} developed a more accurate
code. But since Huang et al.'s code is more convenient for numerical
solution, we will use their code for the current study.

To fit the overall multiband afterglow of GRB 030329, we basically
need a two-component jet (\citealt{ber03}): a narrow jet with a
small opening angle ($\theta_{\rm 0,n}$) and a very large initial
Lorentz factor ($\gamma_{\rm 0,n}$), and a wide jet with a
relatively larger opening angle ($\theta_{\rm 0,w}$) but with a
smaller initial Lorentz factor ($\gamma_{\rm 0,w}$). These two
components are coaxial, and both of them may expand laterally when
they push through the circum-burst medium. The narrow jet can
produce the observed prompt GRB emission and the early afterglow,
while the wide jet can contribute significantly to the late
afterglow. In our localized energy-injection model, the key
ingredient is the late/slow shells that play the role of
energy-injection. We conjecture that these shells should be launched
by the central engine via the same mechanism as the narrow jet
component. The only difference is that they are launched slightly
later, and most importantly, with much lower velocities. They catch
up with the external shocks very lately, so that they could only
manifest as energy-injections and rebrightenings in the afterglow,
but not prompt gamma-ray emission. It is then reasonable to assume
that the half opening angle of these late/slow shells should be
similar to that of the narrow jet, i.e. $\theta_{\rm 0,n}$.

Before encountering the external shock, these late/slow shells move
in the wake of the external shock. Since the circum-burst medium has
been swept-up by the external blastwave, the wake should be very
clean. It means that the movement of the late/slow shells should be
nearly constant. So these shells must be cold and will not expand
laterally. When they finally catch up with the external shock,
energy injection will only occur at the central region of the GRB
remnant, with the involved half opening angle (i.e., the
energy-injection angle) still being $\theta_{\rm 0,n}$. In Figure 1,
we present a schematic illustration of the scenario.

\begin{figure}[h!!!]
\centering
\includegraphics[width=9.0cm, angle=0]{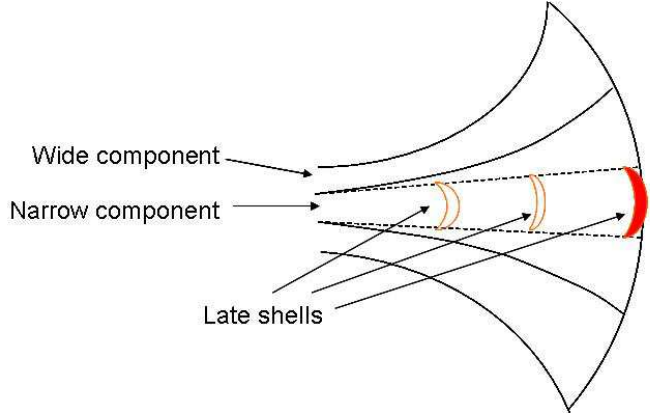}
\caption{A schematic illustration of the localized energy-injection
process. The outer solid curves represent the wide component. The
middle solid curves represent the narrow component. Both the wide
and the narrow components can expand laterally. The inner dashed
straight lines represent the trace of the late/slow shells, which
coast freely and experience no lateral expansion before catching up
with the external shock front. The half opening angle of the
late/slow shells is assumed to equal to that of the narrow jet
component, since they are most likely be of the same origin. }
\label{Fig1}
\end{figure}

\section{Rapidness of the Rebrightening}
\label{sect:Rapidness}

An energy-injection can naturally lead to a rebrightening in the
afterglow. In the case of a localized energy-injection, the
rebrightening should be much steeper as compared to a usual
homogeneous energy-injection. A few parameters may affect the
rapidness of the rebrightening, such as the number density of the
circum-burst medium ($n$), the half opening angle of the late/slow
shells ($\theta_{\rm 0,n}$), the energy fraction of electrons with
respect to protons ($\epsilon_{\rm e}$), the energy fraction of
magnetic field with respect to protons ($\epsilon_{\rm B}^{2}$), the
power-law index of electron distribution function ($p$), etc.
Analytically, the evolution of the radius ($R$) and Lorentz factor
($\gamma$) of the external shock is $R \propto n^{-1/4} t^{1/4}$,
$\gamma \propto n^{-1/8} t^{-3/8}$ (\citealt{hua98}). The angular
timescale, i.e. the approximate duration of the rebrightening, can
then be derived as $\Delta T \approx R_{\rm inj} (1 - \cos
\theta_{\rm 0,n})/c \approx R_{\rm inj} \theta_{\rm 0,n}^2/ 2c
\propto n^{-1/4} \theta_{\rm 0,n}^2 t_{\rm inj}^{1/4}$, where
$R_{\rm inj}$ is the radius of the external shock at the
energy-injection moment of $t_{\rm inj}$. It indicates that the
effect of $\theta_{\rm 0,n}$ on the rapidness of the rebrightening
is most striking, and the parameter $n$ only has a minor effect on
the rebrightening, while the effects of other parameters are very
insignificant.

We have numerically investigated the effects of the involved parameters.
Our numerical results are consistent with the above analysis.
In Fig.~2(a), we illustrate the effect of $\theta_{\rm 0,n}$ on the
afterglow light curve. In these calculations, we take
$\epsilon_{\rm e}$ = 0.1, $\epsilon_{\rm B}^{2}$ = 0.01, $n$ = 1
cm$^{-3}$, $p$ = 1.9, and evaluated $\theta_{\rm 0,n}$ as 0.02,
0.035, and 0.06 respectively. It is quite clear that a small
energy-injection angle can lead to a very rapid rebrightening, and a
large injection angle can make the rebrightening very smooth. It
strongly hints us that we could use the rapidness of the observed
rebrightening to derive the energy-injection angle.

As a comparison, we illustrate in Fig.~2(b) the effect of $n$ on the
afterglow light curve. Again we have taken $\epsilon_{\rm e}$ = 0.1,
$\epsilon_{\rm B}^{2}$ = 0.01, $p$ = 1.9,  and $\theta_{\rm 0,n} =
0.035$, but let $n$ vary between 0.5 cm$^{-3}$ and 2 cm$^{-3}$. We
see that the effect of $n$ on the rapidness of the rebrightening is
much weaker.

In short, we find that the rapidness of the
rebrightening is primarily determined by the energy-injection angle
(i.e. the initial beaming angle). We suggest that we could use the
observed afterglow rebrightenings to measure the beaming angle of
GRBs. This is a novel and hopeful method to constrain the
collimation degree of GRBs, independent to the conventional
jet-break timing method.

\begin{figure}
\includegraphics[width=75mm]{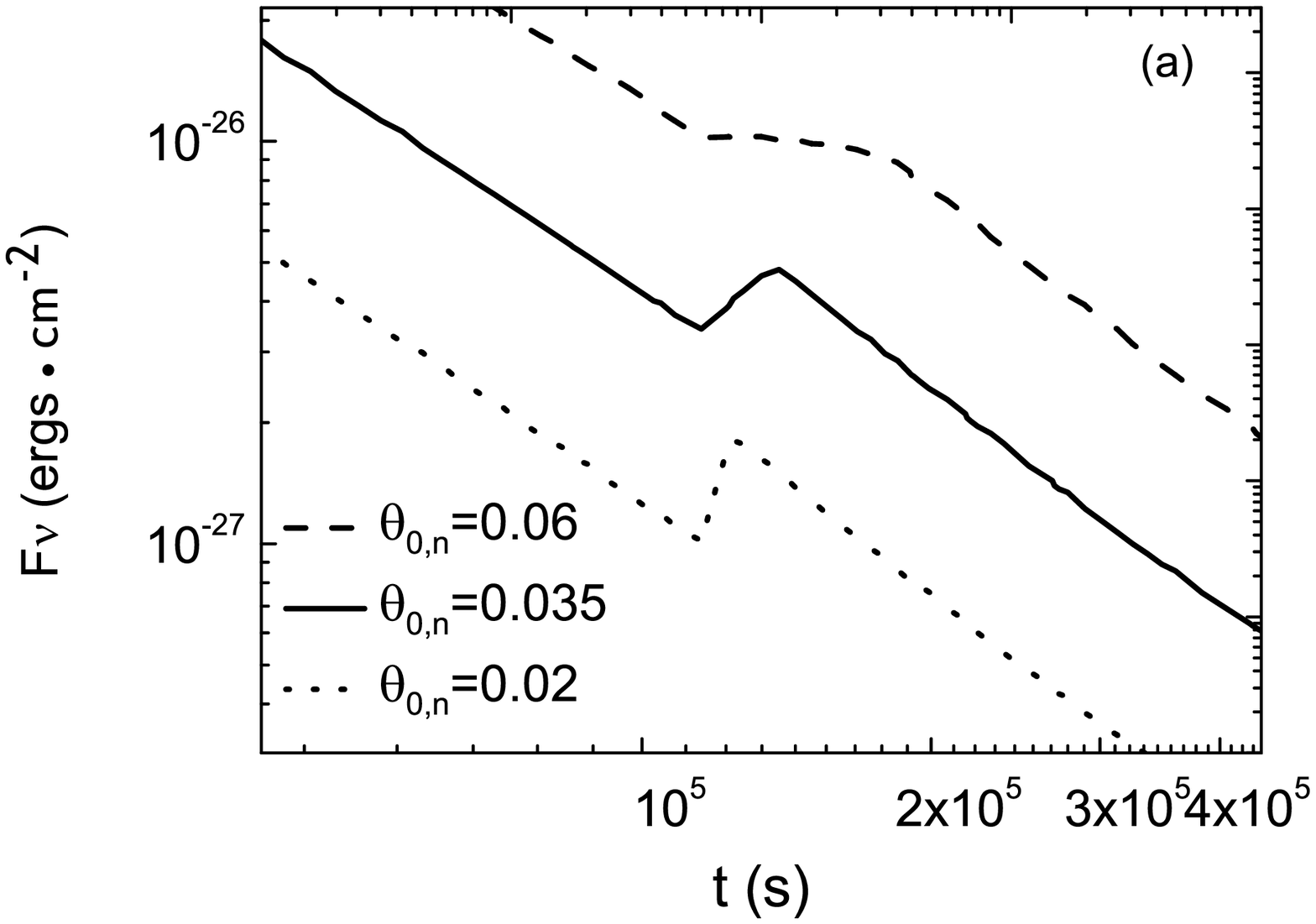}
\includegraphics[width=73mm]{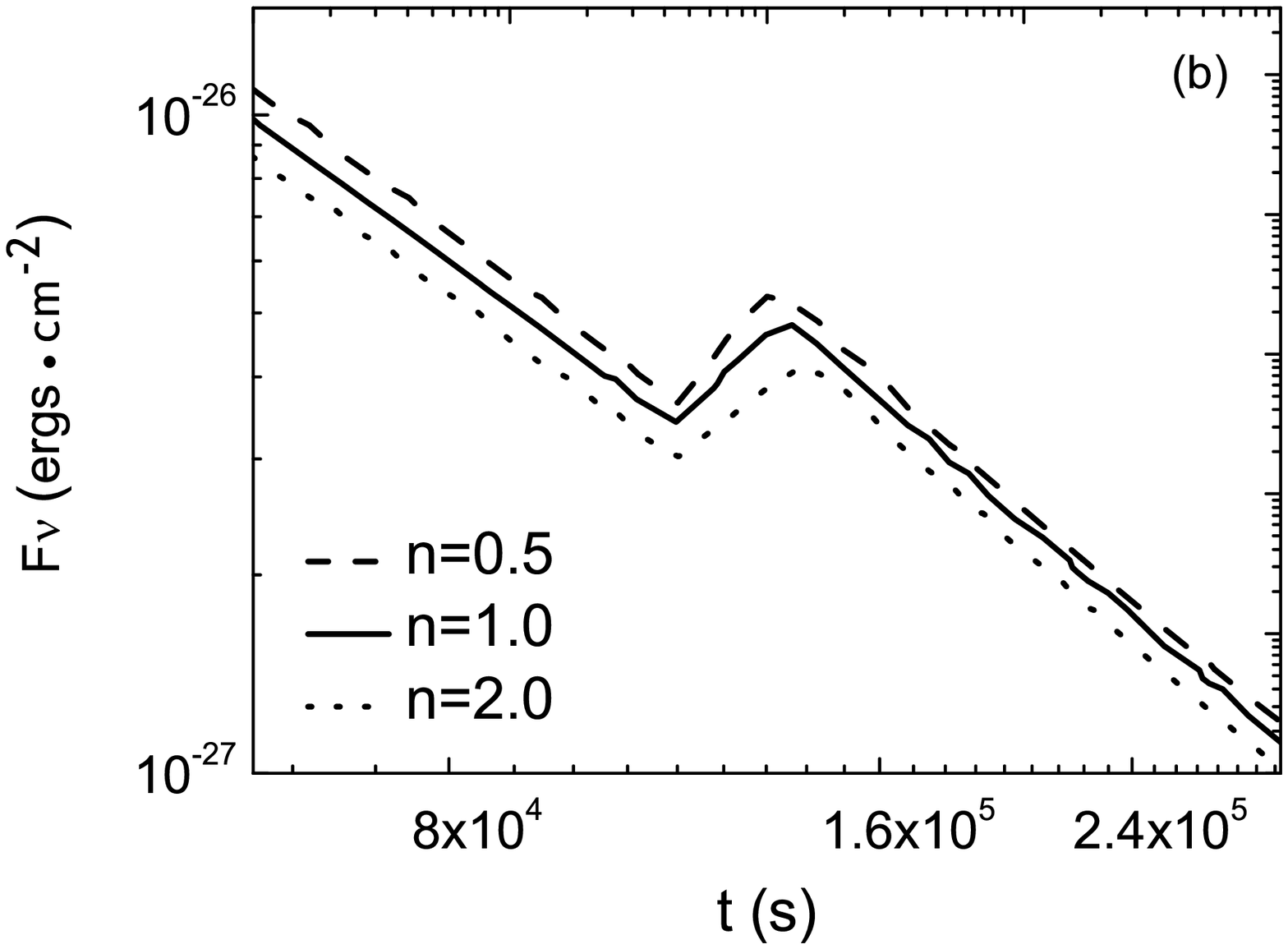}
\caption{The effects of the parameter $\theta_{\rm 0,n}$ (a) and $n$
(b) on the rebrightening caused by a localized energy-injection. In
Panel (a), $\theta_{\rm 0,n}$ is in units of radian, and in Panel
(b), $n$ is in units of cm$^{-3}$. The values of other parameters
are set constant in these calculations and are explained in the main
text of Section 3.} \label{Fig2}
\end{figure}

\section{Application to GRB 030329}
\label{sect:Application}

In this section, we use the localized energy-injection model to
explain the multiband afterglow light curves of GRB 030329. Special
attention will be paid to the remarkable rebrightening at t $\sim$
1.2 $\times$ 10$^{5}$ s. The observed data of optical afterglow are
provided by \cite{lip04}. However, the original data include the
contribution from the host galaxy (\citealt{gor05}) and the
underlying supernova (\citealt{zeh04, zeh05}). They are also
contaminated by the extinction of our Galaxy (\citealt{sch98}). The
pure optical afterglow light curves of GRB 030329 are available only
when these contaminations are corrected for. We will use these pure
afterglow light curves as the final template. The observed date of
radio afterglow can be found in \cite{ber03}. The effect of
synchrotron self-absorption is considered in our calculations.

To fit the overall multiband afterglow of GRB 030329, we basically
need a two-component jet model (\citealt{ber03}). The central narrow
jet component can account for the prompt GRB emission and the early
afterglow, while the wide jet component can account for the late
afterglow. The observed multiple rebrightenings are then produced by
a few energy-injections. In our framework, the energy-injection is
caused by the late/slow shells ejected from the central engine.
These shells should have the same origin as that of those early
shells which generate the prompt GRB emission via internal shocks,
except that they are ejected in a slightly late period and with much
lower velocities. It is then natural to assume that the initial half
opening angle of these shells equal to that of the early shells
(i.e., the narrow jet component). The collision between a late shell
and the external shock is very quick. We can approximate it as an
instant energy-injection process (\citealt{ree98}; \citealt{kum00};
\citealt{sar00}; \citealt{pir04}; \citealt{kin05}).

To fit the observed multiple rebrightenings, we find that at least 5
energy-injections are needed. After correctting for the cosmological
redshift, the injection time (and the amount of energy needed) is
$1.1 \times 10^{4}$ s (0.16 $E_{\rm n,iso}$), $1.0 \times 10^{5}$ s
(0.15 $E_{\rm n,iso}$), $1.8 \times 10^{5}$ s (0.09 $E_{\rm
n,iso}$), $2.2 \times 10^{5}$ s (0.04 $E_{\rm n,iso}$), and $3.6
\times 10^{5}$ s (0.06 $E_{\rm n,iso}$), respectively, where $E_{\rm
n,iso}$ is the isotropic-equivalent kinetic energy of the narrow
component. In Figs. 3 and 4, we illustrate our best fit to the
multiband afterglow of GRB 030329. In these calculations, the
initial half opening angle, the initial Lorentz factor and the
isotropic-equivalent kinetic energy of the wide jet component are
taken as $\theta_{\rm 0,w}$ = 0.2, $\gamma_{\rm 0,w}$ = 50, and
$E_{\rm w,iso}$ = $3.4 \times 10^{53}$ ergs. Correspondingly, the
parameters of the narrow jet component are $\theta_{\rm 0,n}$ =
0.035, $\gamma_{\rm 0,n}$ = 200, and $E_{\rm n,iso}$ = $2.0 \times
10^{53}$ ergs. The common parameters of the two components are
$\epsilon_{\rm e}$ = 0.1, $\epsilon_{\rm B}^{2}$ = 0.01, $n$ = 1
cm$^{-3}$, $p$ = 1.9. The luminosity distance is taken as $D_{\rm
L}$ = 0.8 Gpc. Note that we have assumed a $p$ value smaller than 2.
Although such a small $p$ is not common for other events, it is
still possible in GRBs (\citealt{dc01}).

In Fig. 3, we see that the overall R-band afterglow light curve can
be well explained. Especially, the observed rapid rebrightenings can
be satisfactorily accounted for. Taking the most obvious rebrightening
observed at $\sim 1.2 \times 10^5$ s as an example, this rebrightening is so
rapid that all previous studies failed to present a satisfactory
explanation. For example, in the study by \cite{hua06}, since the
energy was assumed to be homogeneously injected into the whole
external shock, the theoretical rebrightening is then much slower
than observations, and the relative residual is generally on a level
of $\sim$ 20\% at the rebrightening moment. In our current study,
the rapidness problem is satisfactorily resolved. The relative
residual is only about 5\% --- 10\% during the rebrightening
(see the bottom panel of Fig. 4).

Fig. 4 shows that most of the multiband afterglow light curves can
be satisfactorily reproduced. The most obvious deviation appears in
the 4.86 GHz light curve. Our theoretical flux is lower than the
observed intensity after $\sim 2 \times 10^7$ s. Note that at this
late stage, the decay of the observed 4.86 GHz flux density becomes
very slow and the light curve becomes flatter. It is possible that
the contribution from the host galaxy begins to emerge from this
point on. However, it is also possible that the structure of the
ejecta in GRB 030329 were much more complicated than what we have
considered here. In fact, \cite{van05} have argued that additional
component is necessary to account for the excess of the observed
flux at late stages. Finally, radio afterglow is subjected to
scintillation due to scattering of interstellar medium
(\citealt{fra03}). This may also explain part of the deviations,
especially at early stages. Here, since our interest is mainly on
the earlier optical rebrightenings, we suspend from extending our
discussion to more of the above details.

Many other authors have studied the afterglow of GRB 030329
(\citealt{ber03}; \citealt{gra03}; \citealt{hua06};
\citealt{gao10}). It is interesting to note that the derived
parameters vary from one group to another. For example, the
suggested beaming angle of the narrow jet is 0.05 by \cite{hua06}
and \cite{gao10}, 0.07 by \cite{gra03astroph}, and 0.09 by
\cite{ber03}. The suggested beaming angle of the wide jet is 0.3 by
\cite{ber03}, and 0.15 by \cite{hua06}. In our current study, we
take $\theta_{\rm 0,n} = 0.035$ and $\theta_{\rm 0,w} = 0.2$.
Although these two parameters are again different from the above
values, they are not unreasonable, and should be acceptable.
Similarly, our other parameters are also in reasonable ranges.

\begin{figure}[h!!!]
\centering
\includegraphics[width=9.0cm, angle=0]{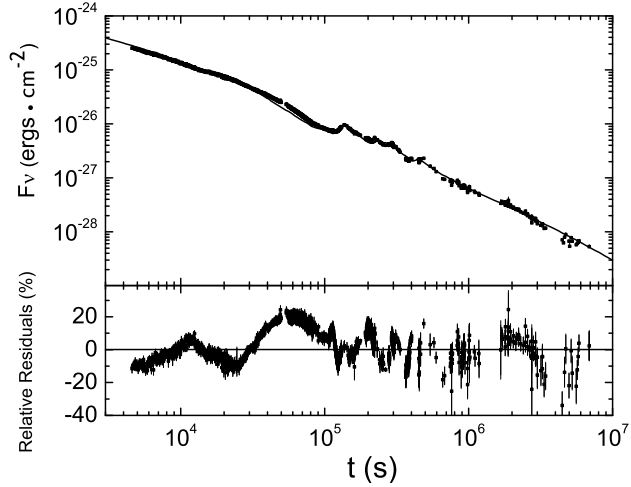}
\caption{Our best fit to the R-band afterglow light curve of GRB
030329, using the localized energy-injection model. Observed data
points correspond to pure afterglow emission (\citealt{lip04}). The
solid line corresponds to our best simulation with five major
energy-injection processes which occur at $1.1 \times 10^{4}$ s,
$1.0 \times 10^{5}$ s, $1.8 \times 10^{5}$ s, $2.2 \times 10^{5}$ s,
and $3.6 \times 10^{5}$ s respectively. The bottom panel gives the
relative residual of the fit. See the main text in Section 4 for
more details of this calculation.} \label{Fig3}
\end{figure}

\begin{figure}
\includegraphics[width=75mm]{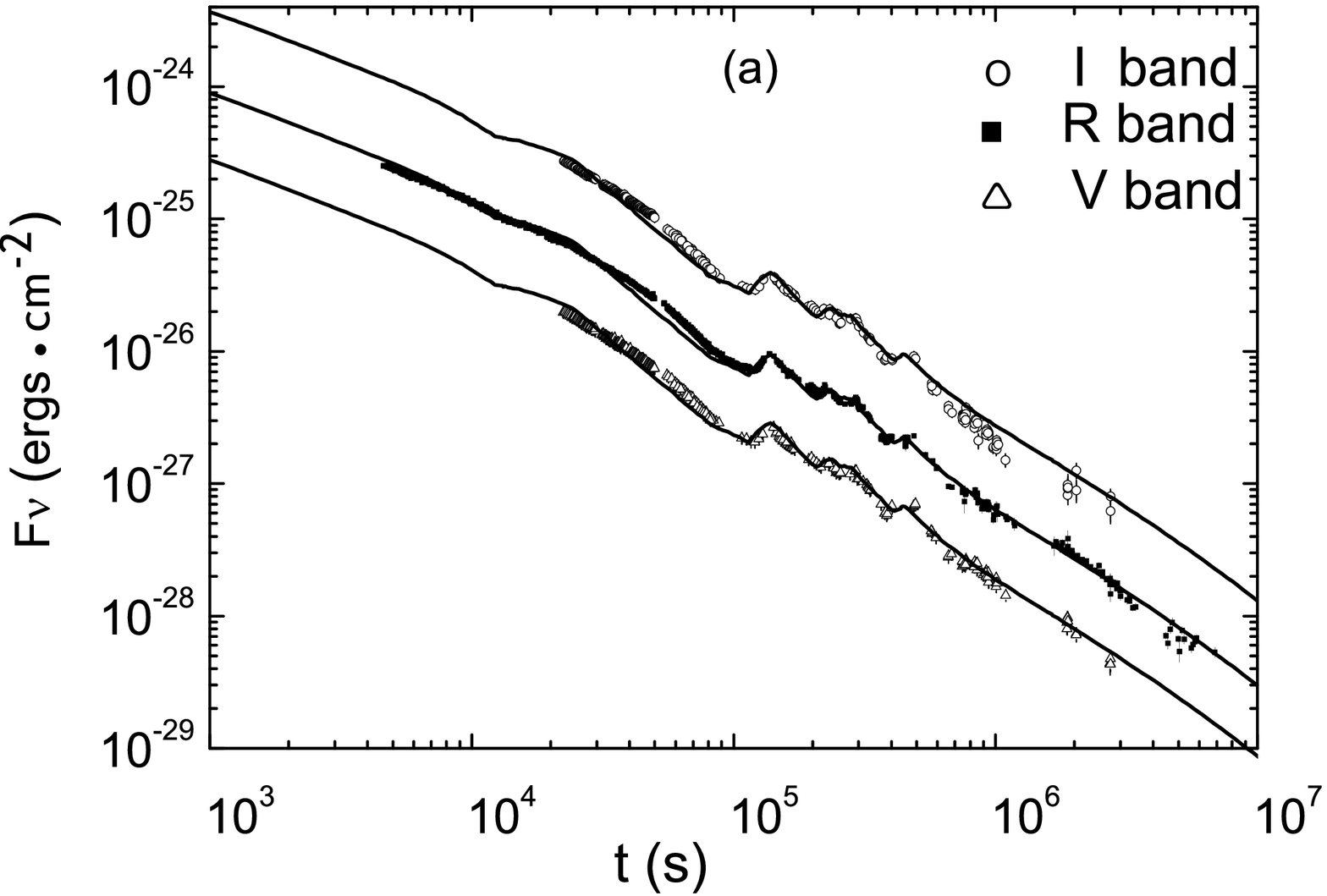}
\includegraphics[width=73mm]{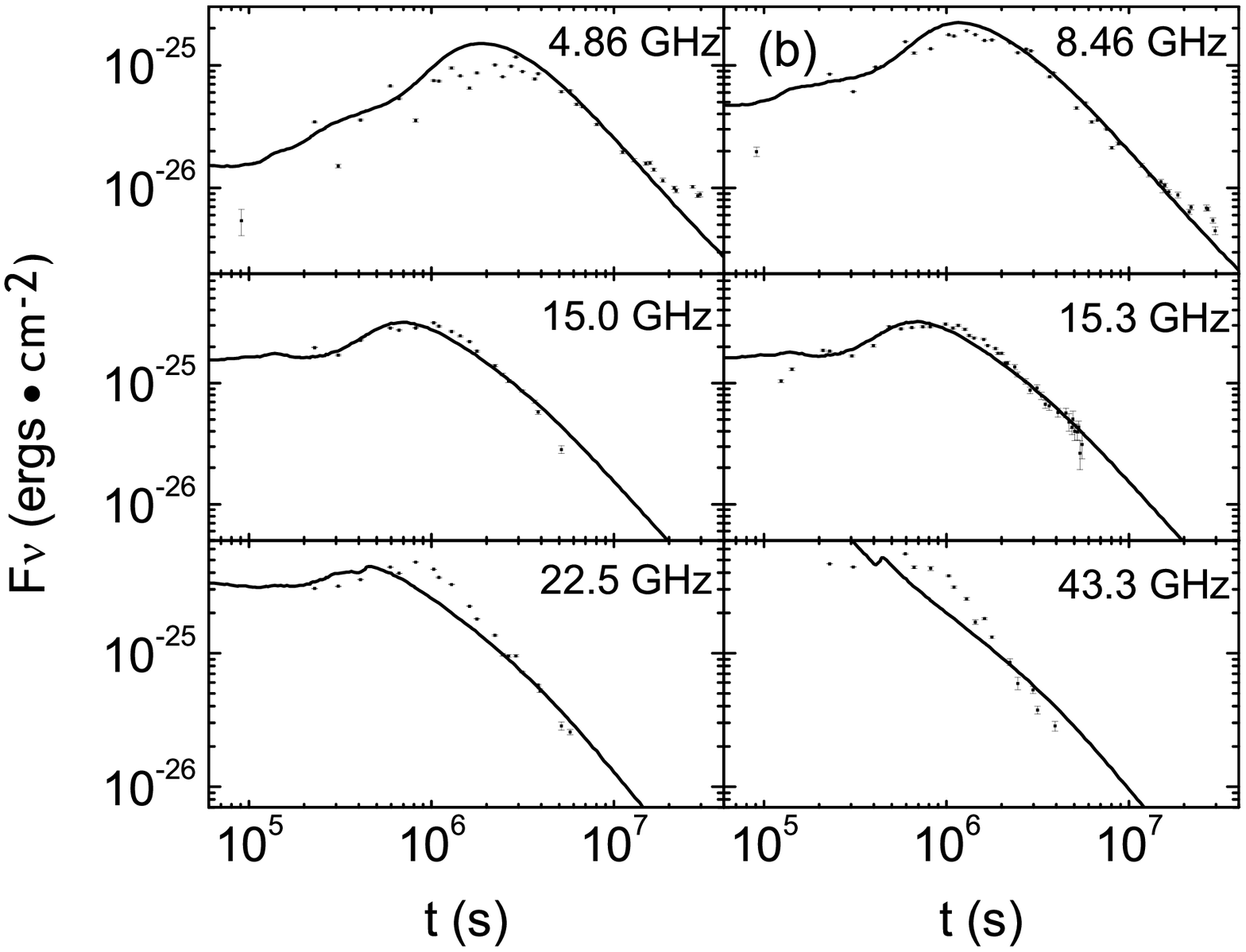}
\caption{Our best fit to the multiband afterglow light curves of GRB
030329 using the localized energy-injection model. Panel (a)
corresponds to optical afterglow. The observed data points are
derived from \cite{lip04}. Panel (b) corresponds to radio afterglow.
The observed data points are taken from \cite{ber03}. Note that in
Panel (a), the I and V band light curves have been shifted by $-1.2$
and $+1.2$ magnitude respectively, for clarity. See the main text in
Section 4 for more details of this calculation.} \label{Fig4}
\end{figure}

\section{Discussion and Conclusion}
\label{sect:Conclusion}

The rapid rebrightening is an interesting feature of the afterglow
of GRB 030329. Conventional homogeneous energy-injection model
cannot account for the rapidness of the observed rebrightenings. In
this study, it is clearly shown that the localized energy-injection
model can satisfactorily resolve the rapidness problem. By using a
two-component jet model, together with 5 localized
energy-injections, we can satisfactorily reproduce the overall and
multiband afterglow light curves of GRB 030329. In our scenario, the
narrow component gives birth to the prompt GRB emission and the
early afterglow, while the wide component contributes significantly
to the late afterglow. The energy-injections are induced by 5 late
shells, which should be ejected by the central engine via the same
or similar mechanism as the narrow jet component, but at a slightly
later period and with much lower velocities. These shells coast
freely in the empty environment after the external shock. They keep
to be cold and do not experience any lateral expansion until they
catch up with the external shock and inject their energies into the
GRB remnant. The energy-injection is a prompt process, and most
importantly, is a localized collision. It can lead to a very rapid
increase in the afterglow brightness, barely subjected to the
smearing of the equal arrival time surface effect.

The initial beaming angle is an important parameter of GRBs. It can
provide important clues for the central engine (\citealt{cl01};
\citealt{zh07}; \citealt{gao10}). According to our best fit, the
energy-injection angle is about 0.035 in the case of GRB 030329. In
our modeling, this angle also equals to the half opening angle of
the narrow jet component. We also showed that the rapidness of the
rebrightening is mainly controlled by the energy-injection angle.
Other parameters, such as $\epsilon_{\rm e}$, $\epsilon_{\rm
B}^{2}$, $p$, and $n$ only have minor or negligible effect. So, we
suggest that the rapidness of the observed rebrightenings can
basically be used to derive a measure of the beaming angle of GRBs.
This is a novel and hopeful method, independent to the conventional
jet-break timing method. Comparing with the jet-break timing method,
the advantage of our method is that it does not rely on the
assumption of the speed of lateral expansion, which itself is quite
unclear currently. The impact of another factor, i.e. the density of
the circum-burst environment, is also very week in our method.
However, the main restriction of our method is that it is applicable
only when rapid rebrightenings due to energy-injections were
observed in the afterglow.

In our modeling of the five observed rebrightenings of GRB 030329,
the beaming angles of the 5 energy-injections are set to be
constant. The fitting results are good. It means that the ejecting
angle of the central engine might not vary significantly over the
active period. However, in the future, it is possible that we could
find new examples which need to vary the energy-injection angle.
Such information will also provide interesting constraints on the
central engine. In our model, we have assumed that the
energy-injection angle equals to the beaming angle of the narrow jet
that produces the prompt GRB. Although this assumption is a natural
hypothesis, it is still possible that in reality they might not
equal, since the real central engine may be much more complicated as
compared with our imagination. Even in this case, the
energy-injection angle derived from observations still can give
useful hints on the characteristics of the central engine.

In our current study, in order to get an acceptable fit to the
overall and multiband afterglow of GRB 030329, we have adopted a
two-component jet geometry. However, note that such a two-component
configuration is not always necessary for localized
energy-injection. For example, in other GRBs, it is quite possible
that the prompt GRB jet may only have one component. This component
can expand laterally during its deceleration, still leaving a
relatively clean environment behind it. So, late shells will still
coast freely after the external shock. They keep to be cold and will
not experience sideways expansion before the final energy-injection.

For GRB 030329, in order to get an acceptable fit to the overall and
multiband afterglow, we have adopted a $p$ value that is less than
2. However, as stated in section 3, $p$ does not influence the
rapidness of the rebrightening. So $p<2$ is not necessary for the
localized energy-injection model. In other GRBs, different $p$ can
be adopted according to observations.

In many GRBs, multiple optical and/or X-ray flares are observed
to be superposed on the early afterglow light curves. The rising and
falling of these flares are even more rapid than that in the case of
GRB 030329. Although the nature of these flares is still largely
unknown, we suggest that useful information on the beaming angle of
these flares can also be derived by fitting the rising and falling
profiles.

\normalem
\begin{acknowledgements}
We thank the anonymous referee for helpful comments and suggestions,
and X. Wang for useful discussion. This work was supported by
the National Natural Science Foundation of China (Grant No.
10625313) and the National Basic Research Program of China (973
Program, Grant 2009CB824800).
\end{acknowledgements}

\label{lastpage}

\end{document}